\documentclass[%
 reprint,
nofootinbib,
 amsmath,amssymb,
 aps,
]{revtex4-2}

\usepackage{graphicx}
\usepackage{dcolumn}
\usepackage{bm}
\usepackage{hyperref}
\usepackage[mathlines]{lineno}
\usepackage{multirow}
\usepackage{tabularx}
\usepackage{tabulary}
\usepackage{adjustbox}
\usepackage{xcolor}
\usepackage{tabularray}
\UseTblrLibrary{booktabs}

\renewcommand{\selectlanguage}[1]{}
\newcommand{\rev}[1]{#1}

\begin{document}

\preprint{APS/123-QED}
\title{Surveying the State of Writing Education in Physics and Astronomy}

\author{Briley L. Lewis}
\email{brileylewis@ucsb.edu}
\affiliation{Department of Physics, University of California, Santa Barbara, Santa Barbara, CA 93108 USA}


\date{\today}

\begin{abstract}
Writing is a critical skill for modern science, enabling collaboration, scientific discourse, public outreach, and more. Accordingly, it is important to consider how physicists and astronomers are trained to write. This study aims to understand the landscape of science writing education, specifically in physics and astronomy, in higher education in the United States. An online survey probing various aspects of their writing training in both undergraduate and graduate school was administered to 515 participants who have obtained training in physics and/or astronomy, or related fields, at the level equal to or beyond upper-division undergraduate study. Humanities and writing requirement courses appear to have a key role in general writing education, while laboratory courses and feedback from mentors are the dominant modes of science writing education in undergraduate and graduate school respectively. There is substantial variation in the quality of writing education in physics and astronomy, often dependent on the student's institution and/or mentor. Some participants also report that their success in disciplinary writing was a result of a solid foundation from K-12 education and/or self-direction towards resources; such reliance on past experiences and student background may contribute to inequality in the field. Many participants also stated a clear desire for more structured writing training to be available in the field. We provide suggestions for how to implement such training to meet the needs of the community identified in the survey.
\end{abstract}

\maketitle

\section{\label{sec:intro}Introduction}

Many students and instructors in STEM have negative attitudes towards writing in science, possibly as a result of the deprioritization of writing in science at multiple levels of education \citep{emerson2019m}. However, it is clearly evident that writing is necessary in many aspects of science, e.g. research papers, grant proposals, graduate school and job applications, and public outreach. Effective communication has the potential to improve not only communication between scientists, making scientific discourse and collaboration more efficient and effective, but also communication with the public, which is necessary for improved scientific literacy and public trust in science \citep{bubela2009science,cook2007reporting}. \rev{Additionally, writing plays a role in the teaching and learning of physics/astronomy. Instructors must be able to write clear, coherent assessments, lecture slides, and course notes, while students must be able to coherently explain their work and answers to show conceptual understanding.}

As a result, writing is widely regarded as a highly desirable skill for careers in STEM, including physics and astronomy (P\&A). Career guidance from the American Astronomical Society (AAS) emphasizes communication skills as part of an astronomy career \citep{aascareers}, and recommendations for undergraduate curriculum from the American Association of Physics Teachers (AAPT) list communicating physics as a key learning outcome for physics students in laboratory courses \citep{Kozminski_2014}. Many instructors say that writing is an important goal of their laboratory courses, and place a high value on students' writing skill development \citep{dounas2018characterizing}. Additionally, the recent Phys21 report from the Joint Task Force on Undergraduate Physics Programs from the American Physical Society, National Science Foundation, and American Association of Physics Teachers highlights the need for instruction focused on communication skills to prepare students for modern careers \citep{heron2019joint}. \rev{Similarly, guidance from Effective Practices for Physics Programs (EP3) emphasizes the need for training in communication skills \citep{ep3comm}.} 
Writing and communication skills are also important for physics and astronomy students who choose to pursue non-academic careers, as these skills are listed as some of the most highly desirable traits in a candidate for employment by the National Association of Colleges and Employers' 2024 Job Outlook report \citep{nace}. Despite the recent rise in the use of artificial intelligence (AI) tools such as ChatGPT, the task of scientific writing still requires a skilled human author, who can be assisted but not replaced by AI tools \citep{mingarelli2025scientific}, maintaining the need for developing writing skills in STEM practitioners. 

Writing and writing instruction are therefore an important aspect of students' professionalization, preparing them for careers in science. Writing is essentially a social practice, defining and perpetuating the norms and activities of a discipline, and helps students build and explore their identities as emerging professionals in a field such as P\&A \citep{blakeslee2007writing,creme2010developing,jerde2004preparing,estrem2015disciplinary}. For example, doctoral thesis writing is key to the development of students' science identities as they make the major transition from student to independent researcher \citep{inouye2019developing, maher2013factors}. Publication is a major part of the social practice of writing in STEM, and students view publication as a transformative process that validates their role in the scientific community, especially female students who are often marginalized in STEM \citep{fankhauser2025writing}. Similarly, writing research papers for publication can be seen broadly as an enculturation task, establishing early career scientists as members of the discipline \citep{florence2004learning}.

Incorporating writing in the science classroom has a number of other benefits as well beyond professionalization and career preparation. Writing tasks that mirror real-world writing scenarios can increase student engagement and buy-in \citep{kiefer2008client}. Popular science writing for non-experts, in particular, can also improve early career scientists' ability to reflect on and contextualize their own research work \citep{pelger2016popular}. Writing-intensive courses and explicit reading comprehension assignments have been shown to improve undergraduates' confidence in their ability to read scientific literature and communicate science \citep{brownell2013writing,lewis2025improving}. Additionally, students co-authoring with faculty members highly improves research skill and grad student outcomes \citep{feldon2016faculty}. Writing about science can also promote reflection on the methods of knowledge production in STEM \citep{keys1999revitalizing}, and it is well established that writing-to-learn (WTL) improves student conceptual understanding in physics \citep{bullock2006building,pulgar2021investigating}. Many students lose marks in physics because they fail to express themselves clearly \citep{farrell2001physics} and some student difficulties with quantum mechanics concepts can be traced to difficulties with language \citep{brookes2007using}, while students who completed writing tasks to explain their ideas performed better on science exams \citep{hand2001sequential}. Explicit writing instruction has also been shown to improve students' ability to communicate to the lay public \citep{moni2007using,brownell2013science}.

Despite the clear benefits of writing pedagogy in STEM, the ways in which scientists learn to write anecdotally seem to be highly varied and often insufficient. Many science writing instructors are disciplinary practitioners (i.e. scientists) without training in writing pedagogy, who have potential to be successful writing teachers but need to explicitly reflect on what they think about writing and how they were taught \citep{adler2022writing}. Additionally, many disciplinary faculty see writing as simply a matter of transferring general writing skills, and as something to be ``left to the professionals'' in writing programs \citep{zhu2004faculty}. In undergraduate physics courses, writing activities are most common in upper-division courses, specifically laboratory activities; these courses often emphasize concepts of precision and clarity in their writing instruction, while overlooking novelty/contextualization and presentation of evidence \citep{carzon2024mapping}. 

In the standard ``apprenticeship'' model of scientific training, students often acquire their skills and knowledge---including writing---from experienced practitioners (i.e. their advisors and mentors), but this model relies heavily on students' existing skills before entering the mentoring relationship, provides feedback that is often too indirect to effectively teach students writing concepts, and can be complicated by the often fraught power dynamics in such a relationship \citep{blakeslee1997activity}. This mode of instruction places students in a precarious position where their writing education and writing process are significantly influenced by power-infused relationships, institutional contexts, and gatekeepers of the discipline, particularly for non-native English speakers \citep{li2006doctoral}. Additionally, WTL practices are not widely implemented despite their demonstrated effectiveness, likely due to a lack of science faculty invested in these practices and the absence of resources to guide interested instructors \citep{reynolds2012writing}. Existing practices in STEM writing pedagogy have been shown to produce students who see writing as an endpoint rather than part of the scientific research process and a prescriptive ``formula'' rather than a reflective task, lack explicit metacognition of written discourse, and are only aware of a narrow range of genres, audiences, and expectations for their writing \citep{yore2002scientists,otfinowski2015writing,fankhauser2021participating}.

Although substantial research exists about writing pedagogy for STEM---broadly defined---there is a limited amount of information specifically for physics and astronomy. Dedicated investigation into writing specifically in P\&A is necessary due to the fact that each discipline within the broad category of STEM has its own unique culture and demographics (e.g. physics has a greater gender imbalance than other scientific disciplines \citep{baram2011quantifying}). Importantly, each discipline also has its own unique \textit{genres} that have a complex relationship to disciplinary culture \citep{golebiowski2002interaction}. Disciplinary genres are quantifiably different not only their content, but also their style \citep{alluqmani2018writing}, and knowledge of genre is linked to perception of knowledge-making practices in each discipline \citep{kuteeva2016graduate}. The features of these genres will differ across disciplines \rev{(and even within \textit{subfields} of P\&A, e.g. theory vs. experiment)}, and these unique features need to be made explicit in teaching \citep{hyland2008genre}--therefore motivating effort into disciplinary writing education. 

This study aims to investigate the current state of disciplinary writing education and training for physicists and astronomers in the United States. This work seeks to illustrate the strengths and the gaps in how physicists and astronomers are taught to write in their scientific discipline, illuminating where to focus future efforts in writing training. In Section \ref{methods}, we describe the survey used, the study population, and how they were recruited, with the survey's results presented in Section \ref{results}. Sections \ref{discussion} and \ref{conclusions} then explore the implications of these results, and offer recommendations for future efforts in writing pedagogy in physics and astronomy.

\section{Methods}\label{methods}

The information in this study was collected via an online survey administered to 515 participants who have obtained training in physics and/or astronomy, or related fields, at the level equal to or beyond upper-division undergraduate study. 

\subsection{Survey Design}\label{survey}

A majority of questions were designed specifically for this study to clearly ask what participants see as the \textit{primary} way they learned to write in each stage of their career, and to report the full range of writing education opportunities available to them and which of those they participated in. Questions included both multiple choice responses with the option to write in an ``Other'' response and free response questions. Survey questions were tested with four scientists of varying career stages to ensure the question text was interpreted as expected \citep{litwin1995measure,taherdoost2016validity}. Demographic information was collected using a combination of commonly-used questions probing gender, age, and ethnicity, plus additional relevant questions related to first-generation student status, native language, and current employment role. Participants were also asked to report their undergraduate and graduate institutions, degree titles, and graduation years, for the purpose of understanding participants' career stages, majors, and institution types (e.g. research institution vs. liberal arts college) to properly contextualize their commentary on their writing education. This survey was granted exemption via UCSB IRB, and participants consented to the survey electronically before accessing the survey questions.

\subsection{Recruitment \& Study Population}\label{studypop}

Our study population consists of 515 participants from the broad group of ``physicists and astronomers'' of most career levels in the U.S., defined via two criteria: participants must have an education level of upper-division undergraduate (i.e. junior or senior) or higher, for a degree in physics, astronomy, astrophysics, or similar, and all or part of that undergraduate and/or graduate education must have taken place in the United States. Eligibility was determined via a set of screening questions at the beginning of the survey. Participants were recruited through broad advertising of the survey, primarily via mailing lists and online forums for relevant professional societies (e.g. American Astronomical Society, American Physical Society) and Physics/Astronomy/Physical Science departments at 2- and 4-year colleges and universities across the country. The survey was also advertised via word of mouth (e.g. people sharing the link in their collaborations) and flyers posted at some relevant institutions and conferences. Participants were allowed to skip any question, and as such, the number of responses for each question varies.

Based on demographic information collected in the survey from 70\% of respondents, women are significantly overrepresented in this sample (approximately 42\%) compared to the population of physicists and astronomers in the U.S., where roughly 20-25\% of physics degrees are earned by women each year according to the American Physical Society \citep{apsdiversity}. A fraction of respondents (7\%) identified as non-native English speakers, which may have an impact on their writing education, as many U.S. colleges and universities have additional writing requirements for international and/or ESL (English as a second language) students. These students also often face additional challenges in their writing education and professional writing in their careers \citep{kranov2009s,hyland2002authority}. Additionally, 19\% of respondents identified as first-generation students; this group similarly faces additional challenges in higher education, including in the writing classroom, as they navigate the expectations of the academy and other obstacles \citep{collier2008paper,stebleton2013breaking}. An analysis of the effects of these demographics on writing experience, however, is outside the scope of this work.

A large variety of career stages and roles are represented in this sample as well: 39\% students, 10\% postdoctoral researchers, 26\% tenure track faculty, 6\% adjunct faculty and lecturers, 10\% academic researchers and staff scientists, and 8\% other (including the roles of: K-12 educator, science communication and/or outreach professional, non-academic researcher, science policy officer, college/university administrator, college/university, support staff, medical resident, post-baccalaureate researcher, software developer, and currently unemployed). Respondents similarly attended a wide variety of colleges--over 200 different institutions across the U.S., from community colleges to research universities--for their undergraduate and graduate education.

\subsection{Analysis}

\rev{A majority of analysis in this study is quite straightforward reporting of numbers/percentages of respondents selecting certain answers in the multiple choice survey. Assuming a population of approximately 25,000 physicists and astronomers in the U.S., the margins of error for our full sample of 350-515 range from $\sim$1-5\%, calculated using a Z value corresponding to a 95\% confidence interval. For the free-response questions, responses were thematically coded by the author or counted based on mentions of key words (if a count of responses in a category was desired), or summarized holistically (if a count was not necessary). Questions regarding reasons respondents did not participate in electives and workshops (at both the undergraduate and graduate level), what feedback on writing was helpful vs. unhelpful, and what was the most valuable aspect of respondents' writing education were summarized by the author to capture the various points brought up by respondents.} 

\rev{Responses to questions regarding what could have been improved in respondents' writing education were summarized holistically as well, although we also counted the number of responses mentioning the words ``class'', ``course'', or ``formal training.'' Responses to questions regarding respondents' emotional responses to feedback were similarly counted based on emotion-related words, as illustrated in Figure \ref{words-feedback}. Only three questions were truly coded: responses relating to the value of elective undergraduate science writing courses and whether their education properly prepared them for a career in P\&A. Coding was performed by the author. These were coded into simple categories of positive vs. negative experiences and yes/mixed feelings/no respectively, instead of detailed thematic codes. These codes relied on clearly defined phrases (e.g. ``yes'', ``yes, but'', ``no'') to avoid introducing bias from interpretation of responses by the coder as much as possible.}

\section{Results}\label{results}

\subsection{Undergraduate Education}

94\% of participants (483/515) answered the question: ``What would you consider to be the \textbf{primary} way in which you generally (i.e. in any genre/discipline) learned to write in your undergraduate education?'' A strong majority of respondents (63\%) identified required general writing courses or writing-intensive humanities courses as their primary mode of writing education in their undergraduate career. A small number (16\%) identified their science and laboratory courses as a primary source of writing instruction. 8\% reported their primary writing education as feedback from mentors, 4\% as elective science writing courses, 1\% as extracurricular workshops, and 1\% as feedback from peers. Nearly half of those who reported something \textit{other} than general writing requirements as their primary method of writing education either did not have such requirements or did not participate in such courses. 6\% of respondents report no explicit writing education in their undergraduate curriculum. 

\begin{figure*}
\centering
    \includegraphics[width=\linewidth]{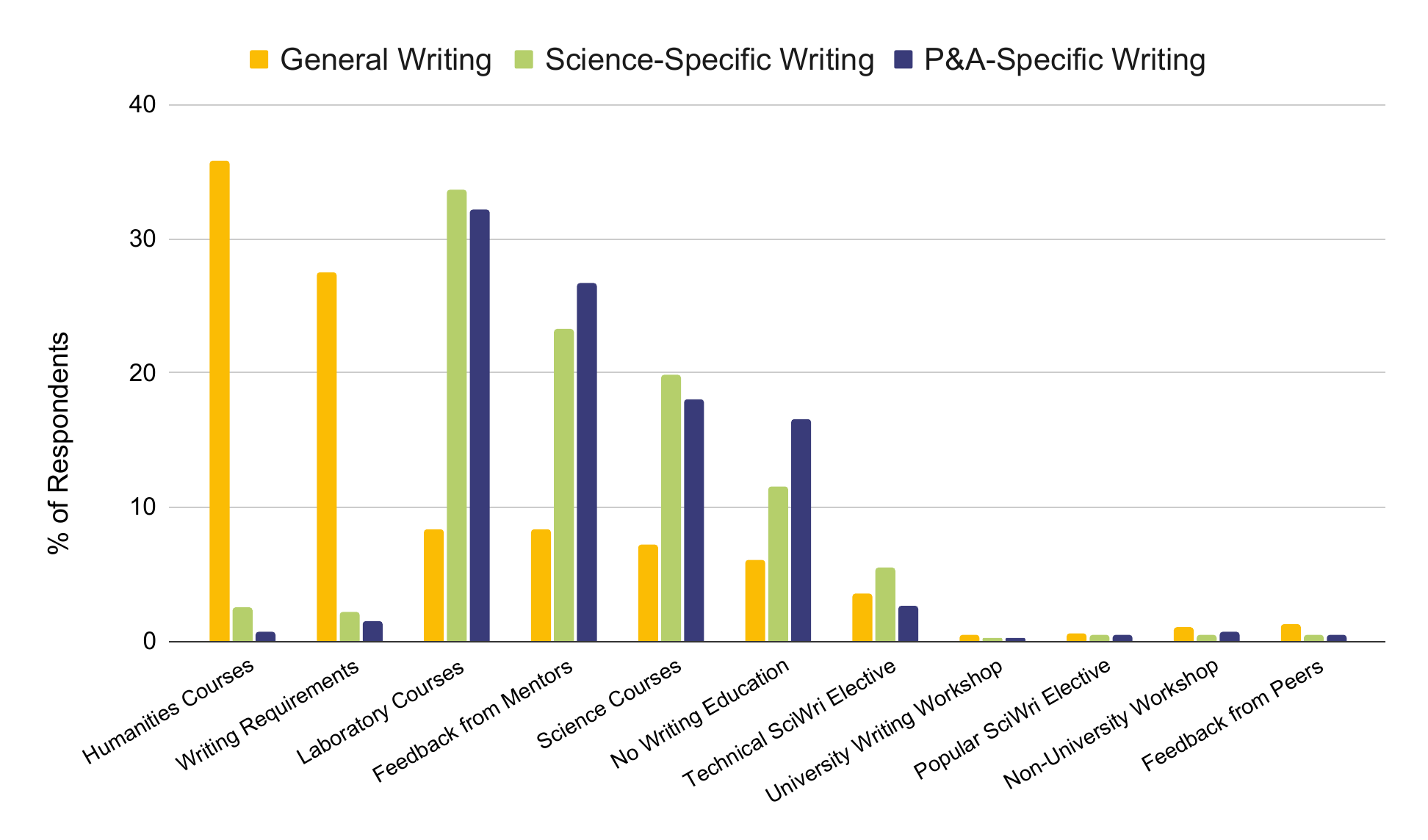}
    \caption{Respondents were asked to report the \textit{primary} way in which they learned how to write in their undergraduate education, both generally (i.e. in any discipline) and specifically focused on science and P\&A. Humanities and writing requirement courses make up the strong majority of general undergraduate writing education, while science- and P\&A-specific writing education are often tackled by laboratory courses or feedback from mentors.}\label{primary-undergrad}
\end{figure*}

91\% of participants (469/515) answered the question: ``What would you consider to be the \textbf{primary} way in which you learned to write specifically about science (any field) in your undergraduate education?'' 34\% report laboratory courses as their primary method of science-specific writing education in their undergraduate careers, while 23\% report feedback from mentors and 20\% report other science courses with writing components. 12\% of respondents report having \textit{no} science-specific writing education, and a small fraction report humanities and other required writing courses (5\%) or elective science writing courses (6\%) as their method of science-specific writing education. Less than 1\% report workshops and conferences as their primary mode of undergraduate science writing education. These results are summarized in Figure \ref{primary-undergrad}.

When asked specifically about writing education in the discipline of physics and astronomy, laboratory courses (32\%, 148/460) and feedback from mentors (27\%) were again the most commonly reported primary modes of instruction. 17\% of respondents report no discipline-specific writing training in their undergraduate education. For those who indicated laboratory courses as their primary mode of instruction for writing in P\&A, nearly half did not have feedback from mentors as part of their undergraduate education.

83\% of participants (426/515) answered the following series of questions about elective course availability: ``Were there any elective courses specifically focused on scientific writing available for you to take during your undergraduate education?'' and ``Were there any elective courses specifically focused on scientific writing in physics and/or astronomy available for you to take during your undergraduate education?'', with a subset also answering the follow-up question ``If yes to either of the above, did you choose to take any of these courses?'' 34\% of respondents report having the option to take a science writing elective course, and only 11\% report having the option for a P\&A-specific writing elective course. Out of those who had a course available to them (254/426), 76\% chose not to participate in these courses. 

Respondents report a variety of reasons for this lack of participation in an open-ended response question on the topic: schedule constraints, a prohibitively heavy load of required courses, considering writing courses unnecessary for their chosen career / not realizing the importance of writing skills, a lack of encouragement to take writing courses, a lack of interest, fear of failure/getting a bad grade, feeling like they had sufficient instruction in their prior education, and not getting into courses with limited space available. Those who \textit{did} participate in elective courses generally report a number of benefits, such as learning \texttt{LaTeX}, preparation for grant writing, practice presenting posters for scientific research, an improved understanding of the scientific peer review and publishing process, and progress towards a senior thesis. Fewer than 10 respondents reported negative experiences in their science writing education, such as a lack of guidance and feedback from instructors, courses perceived as ``useless'' or too rudimentary, and courses offered in different scientific disciplines that didn't feel sufficiently relevant to P\&A. 

84\% of participants (433/515) answered the following question about laboratory courses: ``Did your laboratory courses (if applicable) include a writing component?'' 73\% of respondents say they were expected to write in these courses, but never explicitly taught how to do so. 23\% report both an expectation for writing and writing instruction in these courses, while 3\% report lab courses without any writing and 1.4\% did not take any lab courses. 64\% (260/409) of respondents received feedback from a mentor, with most (70\%) of those advisors being faculty members.

78\% of participants (403/515) answered the following question about extracurricular learning opportunities: ``Did you participate in any workshops (i.e. organized activities outside of formal coursework) geared towards writing skills during your undergraduate education? (Select all that apply)'' A majority of respondents (60\%) did not know of any such opportunities for writing education, whether from their university or beyond, and 28\% knew of opportunities but chose not to participate. Respondents again cite a number of reasons for their lack of participation: nothing specific to their concerns and/or subfield was available, writing was not a priority at the time, poor prior experiences with the groups running such workshops (e.g. campus career centers), a lack of time to participate, a lack of interest, and feeling like their skills and education were already sufficient. 16\% of respondents did choose to participate in extracurricular writing education opportunities available to them, some as part of organized programs, such as the McNair Scholars \citep{bancroft2016mcnair}, the Clare Booth Luce program \citep{daniels2006clare}, the APS Conference for Undergraduate Women in Physics (CUWiP) \citep{poffenberger2020cuwip}, and research experiences for undergraduates (REUs) \citep{follmer2017student}. \rev{ther workshops were  offered by peers, campus writing centers, GRE preparatory courses, and college honors programs.}

\subsection{Graduate Education}

Participants were asked the same series of questions as above for their graduate education, probing their primary method of instruction, the availability of electives, what kind of feedback they received on their writing, and the availability of workshops. A smaller fraction of respondents completed these sections (66-72\%), likely due to the fact that some participants had not yet begun or completed graduate education, e.g. the upper-division undergraduates and first-year graduate students who were included in the study's eligibility criteria.

\begin{figure}
\centering
    \includegraphics[width=\linewidth]{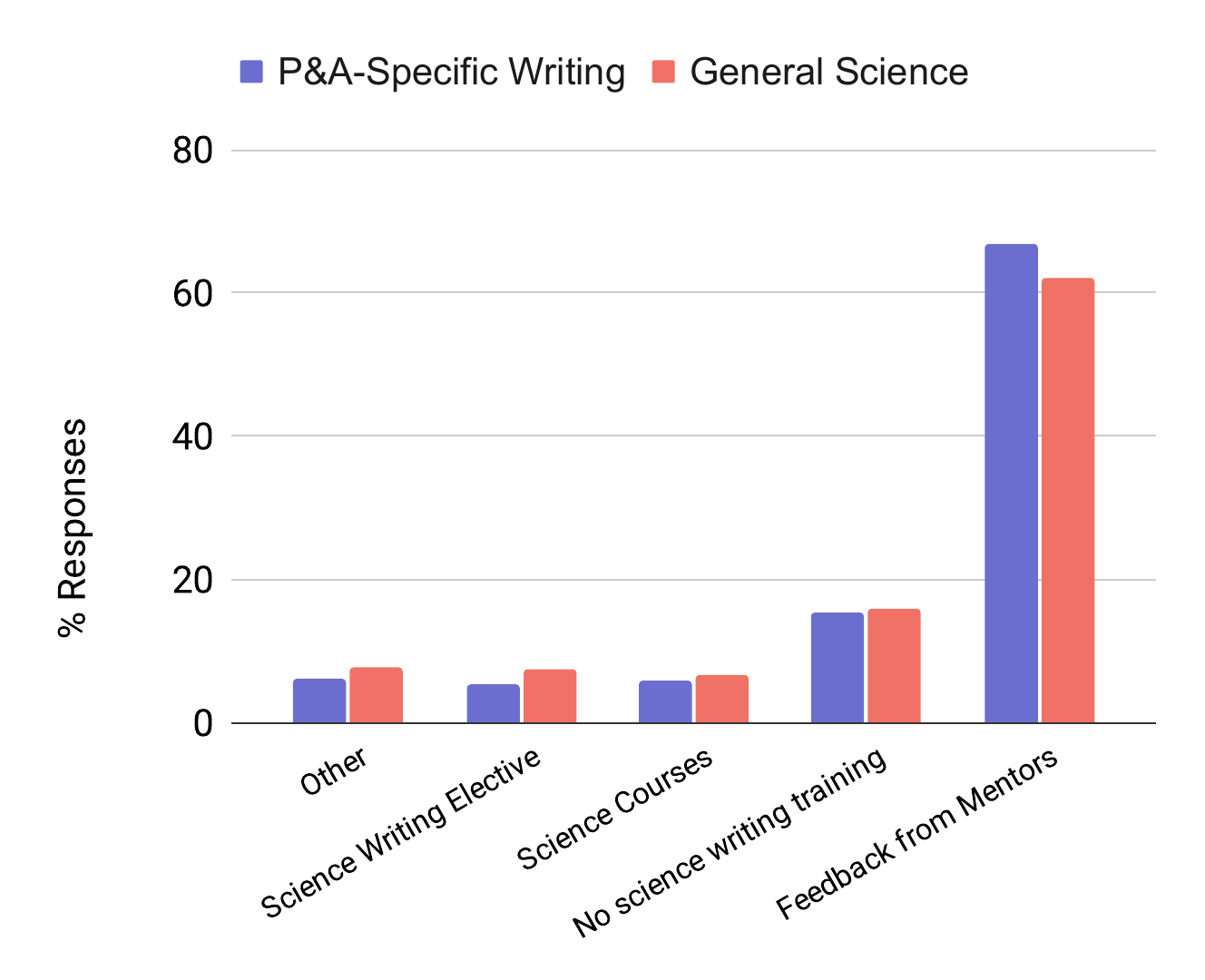}
    \caption{Respondents were asked to report the \textit{primary} way in which they learned how to write about science (both generally, and specific to P\&A) in their graduate education. A strong majority report their primary science writing training was via feedback from mentors, while 16\% report training in some form of course and 16\% report no training at all. Results are similar for general science and P\&A-specific writing.}\label{grad-primary}
\end{figure}

A strong majority of respondents (62\%, 232/373) indicated their primary mode of science writing training in graduate school was feedback from their mentor(s). 16\% reported no science writing training, and 16\% reported their primary mode of science writing training as some form of course instruction (writing-intensive content courses, technical or popular writing electives). Only a few students identified workshops or writing requirement / general composition courses as their primary mode of instruction. Results are very similar for P\&A-specific writing education at the graduate level. These results are summarized in Figure \ref{grad-primary}.

68\% of respondents report \textit{not} having the option to take a general science writing elective course in any discipline during their graduate education. 21\% say they had graduate-level courses available, and 11\% had undergraduate-level courses available. Even fewer had P\&A-specific writing courses available to them. 83.3\% report no such options, while only 11\% had graduate-level courses available and 6\% had undergraduate-level courses available. Out of those who had some type of science writing course available to them (186/359), 79\% chose not to participate in these courses. Reasons for lack of participation were largely similar to the undergraduate level: writing was not a priority, such courses seemed unnecessary or students were already confident in their skills, a lack of interest, a lack of time, existing heavy course loads for required curriculum, and unhelpful/disliked instructors teaching the courses offered. A handful of reasons more specific to the graduate experience were also offered. Some respondents describe the available courses as intended for those with a paper in progress, and they did not have a research paper in the works at the time the course was offered. Others cited a lack of interest in taking additional coursework after their required classes were over, cost for undergraduate courses not included in their program's tuition remission, courses being offered sporadically and the timing never working for them, and other challenges in graduate school that prevented them from taking additional coursework (e.g. travel for research, focus on major exams and milestones).

Nearly all participants (91\%, 314/346) responded yes when asked ``During your graduate education, did you receive feedback on your writing from a research mentor?'' For over 90\% of respondents, these graduate mentors were faculty members. Space was provided for respondents to elaborate on their experience in response to the question: ``If yes, what did you find helpful and/or unhelpful about this feedback?'' A majority of those who elaborated on their responses deemed their mentor's feedback helpful. In particular, it was helpful when the feedback was constructive, iterative, and drew on the mentor's expertise on both the content and the norms of the field. Some, however, reported feedback that ranged from unhelpful to actually harmful. This type of feedback was generally characterized by minimal commentary, slow response times, negative attitudes, and rewriting the student's work without collaboration. 

\begin{figure}
    \centering
    \includegraphics[width=\linewidth]{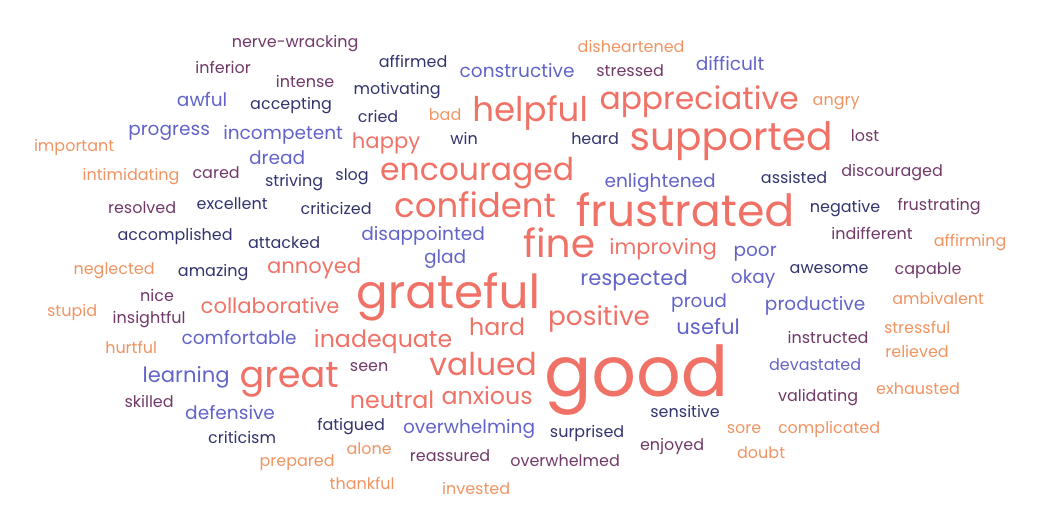}
    \caption{Words related to feelings present in the responses to the question ``If yes, how did this feedback make you feel?'' about feedback from a mentor during graduate education. Words that appear more frequently in the responses are shown in larger font size. The top five words that appear are good (22 occurrences), grateful (13), frustrated (12), fine (10), and supported (10).}
    \label{words-feedback}
\end{figure}

Participants were also asked about their emotional response to the feedback from their mentor(s) with the question: ``If yes, how did this feedback make you feel?'' Responses here were more mixed, where some who reported helpful feedback also reported difficult emotions. Positive responses included words such as: grateful, helpful, appreciative, motivating, collaborative, confident, encouraged, enlightened, affirmed, supported, reassured, valued, productive, supported, and respected. Negative responses included words like: devastating, annoyed, anxious, incompetent, attacked, disappointed, inadequate, awful, dread, criticized, stupid, defensive, frustrated, discouraged, overwhelming, difficult, exhausted, and disheartened. A word cloud of feelings-related words from qualitative responses to this question is shown in Figure \ref{words-feedback}; note that some words were adjusted for tense or form while retaining meaning (e.g. frustrated vs. frustrating).

Similar to the undergraduate-level, a majority of respondents (56\%, 192/343) did not know of any extracurricular opportunities (e.g. conferences, workshops) available to improve their writing skills. 30\% chose not to participate, and 18\% did participate in some form of extracurricular writing education. Again, reasons provided for the lack of participation include: lack of prioritization of writing, lack of specificity to student needs, feeling sufficiently prepared, time constraints, lack of interest, and lack of emphasis by mentors/peers. Some extracurricular activities named as useful that respondents participated in include programming from university writing centers, ComSciCon (a science communication workshop for graduate students) \citep{sanders2018comscicon,mostaghim2023comsciconcan}, American Astronomical Society meetings, Astrobites \citep{khullar2019astrobites}, events from the National Center for Faculty Development \& Diversity (NCFDD) \citep{hanasono2021setting}, and community-organized peer workshops geared towards fellowship applications (e.g. NASA FINESST, NSF GRFP).

\subsection{Overall Attitudes about Writing Education in P\&A}

\begin{figure*}
    \centering
    \includegraphics[width=0.65\linewidth]{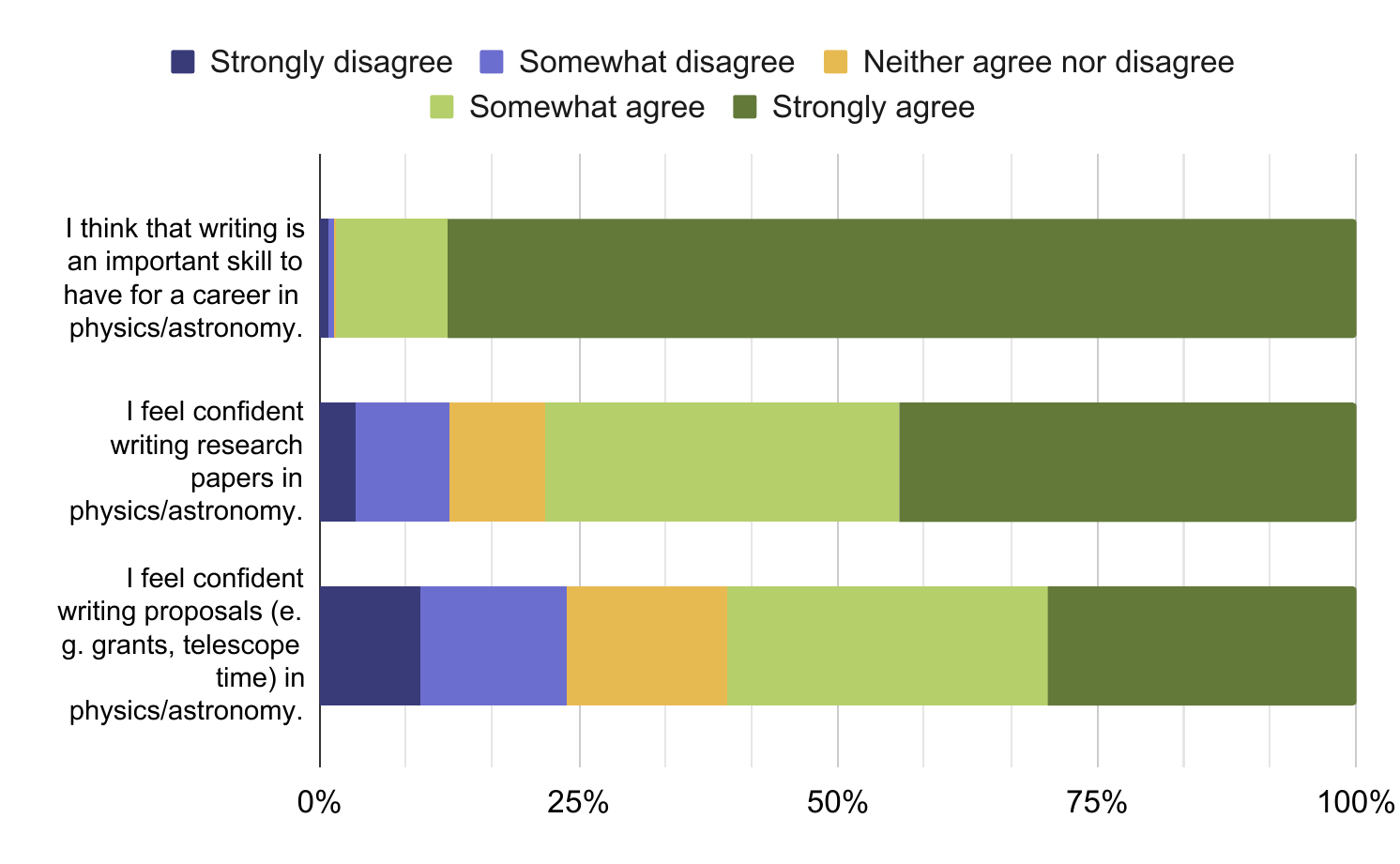}
    \caption{A strong majority of respondents agree that writing is an important skill for physics and astronomy, and a smaller (but still strong) majority are confident in their writing skills.}
    \label{attitudes-bar}
\end{figure*}

72\% of participants (372/515) answered three Likert scale questions about their attitudes towards writing in physics and astronomy, on a five point scale from strongly disagree to strongly agree. Nearly all respondents (98\%) agreed to some extent (somewhat or strongly agree) that writing is an important skill to have for a career in P\&A. A smaller fraction, but still a majority, feel at least somewhat confident writing research papers (78\%) and proposals (61\%). These results are shown in Figure \ref{attitudes-bar}.

Participants were also asked a variety of free response questions about their writing education. When asked about the most valuable aspect of their writing education for preparation for a career in P\&A, respondents most often cited practice writing in real contexts (e.g. research papers, proposals, fellowships, theses). Specific, constructive, iterative feedback was also a common theme in what was valuable. When asked what could have been improved in their writing education, over 60\% (175/285) of respondents mentioned a course/class or other formal training of some kind. These stated desires for courses were accompanied by several descriptors of what would make such training valuable: targeted towards P\&A, involving writing experts in addition to astronomy experts, providing opportunities for both authentic practice and lower-stakes practice, explicit instruction on genre, inclusion of both technical writing and writing for the public, and some kind of requirement or incentive for participation.

When asked if their education properly prepared them for the writing expected in this career, 46\% (145/316) responded affirmatively, 17\% had mixed feelings, and 33\% said their writing education was insufficient. Responses including words like ``yes'' were counted as affirmative, ``no'' for negative, and ``yes, but...'' or similar contradictory messaging as mixed feelings. For those responses categorized as mixed feelings, they can be grouped into six major themes: (1) I made it through adequately, but my education could have been much better; (2) one level of my education was helpful/sufficient (e.g. undergrad), but another (e.g. grad school) was not; (3) my education was enough for research papers, but not other genres like proposals; (4) I feel prepared only because I came in with a solid foundation in writing from K-12; (5) I feel prepared only because I have a natural affinity for writing; or (6) I feel prepared only because I sought out opportunities beyond those required or obviously available.

\section{Discussion}\label{discussion}

Physicists and astronomers in this survey clearly agree on the importance of writing for careers in these fields, even if many of them are not yet fully confident in their own writing. Respondents' writing education as reported is typically comprised of general writing skills developed in undergraduate humanities and writing requirement courses, some science-specific training in undergraduate laboratory courses, and feedback from mentors to refine their writing in the discipline during graduate school. However, nearly half of respondents were not completely satisfied with their writing education, feeling in some way unprepared for the writing tasks expected of them in their careers. Additionally, over half of respondents expressed that a writing course or other formal training would have improved their writing education. There is clearly a need in the community for better preparation for the substantial writing tasks inherent in a career in physics and astronomy, particularly for future researchers who will be expected to publish scientific papers and submit grant proposals.

In particular, it seems to be an issue that writing education is not consistently and reliably available, and depends on what institution a student attends or who they have as a mentor. There are many excellent opportunties and experiences described by respondents who had a supportive mentor who they saw as a ``good writer'' or whose institutions had writing practice infused throughout their undergraduate curricula, including in science. However, there were also many who reported extremely negative experiences. For example, when asked if their writing education prepared them for their careers in physics and astronomy, some respondents left the following comments:
\begin{itemize}
    \item ``Oh my god, NO!  I've been taught primarily by advice on Reddit and insults from my advisor.  I have no idea what I'm doing!'' 
    \item ``My ability to write grants is very minimal, and as a result has been an impediment.''
    \item ``My educational path completely failed to prepare me for a career that lives or dies on the ability to write well.  What I now know is essentially all self taught.''
\end{itemize}
Feedback from mentors was listed as a major contributor to science writing education at multiple educational levels, but suffers from similar extremes in quality of education. When asked about the helpfulness of this feedback and how the feedback made students \textit{feel}, responses included some particularly harmful experiences. One respondent, for example, told a story of a required class with no guidance that they described as ``humiliating, taught me nothing, and made me want to drop out of college.'' Another said their mentor's feedback was ``Awful.  Absolutely awful.  I cried during the meetings.'' Stories like these emphasize the potential for major impacts--both positive and negative--by mentors in a students' education in the field, including with writing. Although there is a lot of great work being done to support students in their writing, it is concerning that some of the negative experiences are so detrimental. 

There is unfortunately no one clear solution to this problem of ``bad apples'' in the teachers and/or mentors a student may encounter, and in fact, a good deal of research has explored the effects of such experiences in other aspects of mentoring beyond writing education \citep{deitte2024good,tuma2021dark}. This work adds to the existing picture of how negative mentoring experiences can impact students' careers, suggesting that they can prevent students from honing the critical skill of writing that will enable them to successfully participate in scientific publishing, and/or they can instill a sense of insecurity and insufficiency in students with respect to their writing skills. 

For some respondents, their mentors were not outright hostile, but were not particularly helpful in their feedback on writing. One respondent, now in the role of mentor instead of mentee, commented on the difficulty of this aspect of the role, saying ``I spent the first few years of my faculty job learning how to \textit{teach} writing to grad students, because being a strong writer is insufficient for faculty life -- you need to be able to teach others to be strong writers.'' It appears that faculty and other mentors may benefit from additional guidance on how to serve as not only writers, but writing instructors, especially given how prevalent mentor feedback is as a mode of writing education. For those working in writing programs as their main focus, this is likely not a surprise; teaching writing is a skill that needs to be learned and developed, and is an expertise in its own right. Disciplinary writing, however, shouldn't be outsourced entirely to those without disciplinary expertise, such as general writing programs staff. There is substantial value in dialogue between students and experienced practitioners (i.e. more senior members of their field) in a disciplinary genre \citep{zhu2004faculty,goldschmidt2014teaching}. Existing works have repeatedly called for further collaboration between writing and disciplinary experts \citep{emerson2019m,harding2020revising,adler2022writing}. This sentiment is even echoed by some respondents in the survey who explicitly mentioned the value of writing professionals and expressed a desire for courses taught by disciplinary faculty in collaboration with experts from a writing program/department.

It is also worth noting the following recurring sentiment in respondents' commentary: their preparation for writing in their careers depended on a solid foundation in writing from their K-12 education, a natural affinity for writing, and/or a self-directed effort to seek out additional writing education opportunities beyond those required or obviously available. For example, in response to a question about if their education prepared them to write, participants said:
\begin{itemize}
    \item ``I do feel very well-equipped at this point in my career (postdoc) for writing in my field, but on the whole I feel like most of my learning towards that end was unstructured/self-directed reflection in response to spending a lot of time writing (because this career involves a lot of writing!). So overall, I would say no, the actual curriculum did not equip me in this, and while I definitely learned a lot along specific axes from courses/workshops/mentor feedback, most of my current writing acumen can be traced back to 1) a head start, and 2) experience.''
    \item ``My education did, but only due to my privileged access to research mentors and not necessarily my formal training.''
    \item ``I was a talented writer when I was little, and I think I've just been running on my natural skills and my interest in being able to teach and express ideas clearly.''
    \item ``I think that what has been the most valuable is the opportunities I sought out (deciding to write my own telescope proposals, apply for fellowships, etc) so I guess within that process I have learned a lot but I'm not sure if every student who goes through this program gets those same opportunities.''
\end{itemize}
There is an inequality inherent in relying on such circumstances for preparation for success in writing in P\&A, as students from lower socioeconomic statuses often have less access to quality K-12 education \citep{reardon2019geography,hasan2019digitization} and first-generation students face additional challenges in navigating the academy and finding resources \citep{aruguete2017recognizing,collier2008paper,stebleton2013breaking}. 

Although many issues in the status quo for writing education in P\&A were identified in this survey, the positive experiences described can point us towards paths forwards to improve writing education in our field. At the undergraduate level, laboratory courses stand out as a clear place to incorporate writing pedagogy into an already packed course load for physics and astronomy majors; they are widely taken (only 1.4\% reported \textit{not} taking a lab course), and writing is often already an expectation in these courses. For respondents who did not have general writing requirement courses, other courses--such as laboratory courses with writing involved--seem to have at least partially filled that gap. Similarly, for undergraduates without the opportunity to receive feedback from a mentor, laboratory courses could fill that crucial role of science-specific writing training. 

At multiple levels, courses and workshops about general writing skills are usually available, but what students actually need is instruction that is specific to P\&A and the writing tasks that accompany careers in this field (e.g. proposals, fellowship applications, articles for a specific journal). This training should incorporate: authentic tasks (especially if able to piggyback on an existing task the student has, such as writing a research paper or thesis), low-stakes practice opportunities, discussion of genre and writing for both technical and popular audiences, guidance from both writing experts and disciplinary experts, and feedback that is constructive, iterative, and from someone with expertise in the genre and content at hand, based on commentary in this survey. Many of the suggestions brought up by survey participants echo best practices known from disciplinary education research, such as the usefulness of regular practice \citep{kellogg2009training}, iterative feedback \citep{vardi2012impact}, authentic tasks in the classroom \citep{kiefer2008client}.

Survey commentary also suggests that it would also be helpful to have these trainings offered regularly (yearly or more frequently if possible) to provide opportunities for students to participate when their schedules allow and when it is the right time for them in their careers. Training needs to also be well-advertised, and faculty and other mentors are responsible for encouraging students to pursue such training and devote time to honing their writing skills. Ideally, some form of writing training would be required or incentivized to encourage participation of those who are not self-motivated. Similarly, it would be beneficial for the culture of the field to shift towards not only valuing writing, but encouraging work on these skills as a critical part of a research career, just like mathematics and coding. 

\section{Conclusions \& Recommendations}\label{conclusions}

Overall, writing appears to be highly valued in the P\&A community, but the education offered is only sufficient for approximately half of those surveyed in the field. 
From this survey of 515 U.S. physicists and astronomers, we find the following:
\begin{enumerate}
    \item Writing education for P\&A typically begins with humanities courses and writing requirements (e.g. first year composition), which ideally develop students' foundational writing skills. \rev{Beyond this, many are left to develop discipline-specific writing skills informally (i.e. outside of the classroom).}
    \item Laboratory courses are currently the most common place for undergraduate science writing education in P\&A, and are therefore an obvious first target for incorporating and/or improving writing instruction into existing curricula as writing tasks are already expected of students in many lab courses.
    \item Feedback from mentors is the most common form of writing instruction in P\&A-specific contexts, particularly in graduate school. The quality of such feedback, however, is variable and the worst cases are particularly detrimental to students.
    \item Successful preparation for meeting the expectations of writing in P\&A too often depends on self-direction towards resources, choice of mentor or institution choice, and/or K-12 writing foundations, possibly creating inequality in writing education (and, therefore, success more generally) in the field. 
    \item Reasons for not participating in available writing training often include a lack of a sense of importance of writing, a lack of encouragement from faculty/mentors/peers, and a lack of time.
    \item The P\&A community highly values the skill of writing \rev{and sees it as an essential skill for success in a P\&A career}, and wants more opportunities for structured writing education that incorporate helpful content and practices.
\end{enumerate}

\rev{As a result, we strongly argue for the inclusion of structured writing education specifically relevant to the field in undergraduate and/or graduate physics and astronomy programs. To this end, we} propose several recommendations for future writing pedagogy efforts in P\&A:
\begin{enumerate}
    \item Instructors of laboratory courses---especially those that expect students to write (e.g. lab reports, mock papers)---should take that opportunity to provide some explicit instruction and feedback on student writing. \rev{However, many lab instructors are not currently trained to provide such writing education. Accordingly, efforts to provide lab instructors with writing pedagogy skills may have a huge impact on student writing education in P\&A.}
    \item Mentors should be better prepared with resources on not only how to write well themselves, but how to provide constructive feedback on writing and teach writing to their students. Further work is necessary to understand the needs of mentors as writing instructors, and to develop materials to serve these needs.
    \item Resources for writing pedagogy in P\&A should be made publicly available across institutions and advertised well, increasing access for students at institutions without existing writing training or with ineffective mentors for writing.
    \item Given that two major reasons for not participating in writing training are a lack of time and a lack of awareness by students, faculty and other mentors should consider how to incorporate writing pedagogy into existing curricula and encourage students to take available opportunities at both the undergraduate and graduate levels.
    \item Writing training for P\&A should ideally have the following attributes identified as helpful and desired by the community:
    \begin{itemize}
        \item Specifically focused on physics and astronomy
        \item Authentic writing tasks, especially if able to work with a task a student is already working on (e.g. a research paper in progress)
        \item Low-stakes writing practice
        \item Discussion of genre, for both technical and popular/science communication audiences
        \item Guidance from both writing experts and disciplinary experts in collaboration
        \item Feedback that is constructive, prompt, and iterative, ideally from someone with expertise in the genre and content at hand
        \item Regularly offered (i.e. yearly or more frequent if possible)
        \item Well-advertised and encouraged by the community, especially faculty and other mentors
        \item Required or incentivized in some way
    \end{itemize}
\end{enumerate}

It is worth noting that this work does not significantly explore writing education beyond graduate school, but learning takes place throughout a career in physics and astronomy; accordingly, there may be additional opportunities and needs for writing training at other career stages. \rev{This work also does not explore generative AI tools such as ChatGPT, as their use was only mentioned by three respondents (one to highlight the usefulness of proofreading tools like Grammarly, and two to condemn the use of AI) and the use of AI in the writing classroom is still highly uncertain, underexplored, and debated.} However, with this survey information, there are now a number of \rev{clearly identified and currently feasible} paths forward to improve writing education in physics and astronomy at the undergraduate and graduate levels, making the preparation for this critical skill more equal, accessible, and effective for all students.

\begin{acknowledgments}
B.L.L. acknowledges support from the National Science Foundation Astronomy \& Astrophysics Postdoctoral Fellowship under Award No. 2401654. Any opinions, findings, and conclusions or recommendations expressed in this material are those of the author(s)and do not necessarily reflect the views of the National Science Foundation. 

This study was conducted with a UCSB IRB Exemption Certificate IRB\#1-25-0196. The authors would like to thank K. Supriya (UCLA), Graham Read (UIC), and Sarah Roberts (UCSB) for their feedback on the survey's initial design, and Laurel Westrup for her mentorship in UCLA's Writing in the Disciplines course, where the idea for this project originated. Thank you to Anavi Uppal (UCSC) for her proofread of this manuscript. We also thank all anonymous participants for their time and effort in responding to this survey.
\end{acknowledgments}

\bibliography{bib}

\end{document}